\documentclass[10pt,journal,twocolumn]{IEEEtran}
\usepackage{wrapfig,booktabs,fancyhdr,amsmath,amsfonts,tabularx,numprint}
\usepackage{cite,bm,bbm,amssymb,amsthm,url,multirow,times,enumitem,comment}
\usepackage{mathtools,siunitx,balance,tikz,adjustbox,graphicx,array}
\usepackage[font=footnotesize]{subcaption}
\usepackage[linesnumbered,ruled]{algorithm2e}
\usepackage[colorlinks=true, linkcolor=black, citecolor=blue, urlcolor=blue]{hyperref}
\usepackage{float}
\usepackage{dsfont}

\newcommand{\vb}{\boldsymbol}

\newcommand{\vbt}[1]{\tilde{\boldsymbol{#1}}}

\newcommand{\tr}{\mathsf{T}}

\newcommand{\ba}{\begin{array}}
\newcommand{\ea}{\end{array}}

\newcommand{\E}[1]{\mathbb{E}\left[ #1 \right]}

\DeclareMathAlphabet{\mathpzc}{OT1}{pzc}{m}{it}

\DeclareMathOperator*{\argmax}{\arg\!\max}

\begin{document}

	\title{Single-Satellite-Based Geolocation of Broadcast GNSS Spoofers from Low Earth Orbit }
	\author{\normalsize Zachary L. Clements\IEEEauthorrefmark{1}, Patrick B. Ellis\IEEEauthorrefmark{2}, Iain Goodridge\IEEEauthorrefmark{2}, Matthew J. Murrian\IEEEauthorrefmark{2}, Mark L. Psiaki\IEEEauthorrefmark{3}, and Todd E. Humphreys\IEEEauthorrefmark{1} \\
		\emph{\IEEEauthorrefmark{1}Department of Aerospace Engineering and Engineering Mechanics, The University of Texas at Austin} \\ 
		\emph{\IEEEauthorrefmark{2}Spire Global} \\
		\emph{\IEEEauthorrefmark{3}Department of Aerospace and Ocean Engineering, Virginia Tech} 
}
\maketitle
	
\begin{abstract}

This paper presents an analysis and experimental
demonstration of single-satellite single-pass geolocation of a terrestrial
broadcast global navigation satellite system (GNSS) spoofer from low Earth
orbit (LEO).  The proliferation of LEO-based GNSS receivers offers the prospect
of unprecedented spectrum awareness, enabling persistent GNSS interference
detection and geolocation.  Accurate LEO-based single-receiver emitter
geolocation is possible when a range-rate time history can be extracted for the
emitter.  This paper presents a technique crafted specifically for
indiscriminate broadcast-type GNSS spoofing signals.  Furthermore, it explores
how unmodeled oscillator instability and worst-case spoofer-introduced signal
variations degrade the geolocation estimate.  The proposed geolocation
technique is validated by a controlled experiment, in partnership with Spire
Global, in which a LEO-based receiver captures broadcast GNSS spoofing signals
transmitted from a known ground station on a non-GNSS frequency band.

\end{abstract}

	
	\newif\ifpreprint
	\preprinttrue
	
	\ifpreprint
	
	\pagestyle{plain}
	\thispagestyle{fancy}  
	\fancyhf{} 
	\renewcommand{\headrulewidth}{0pt}
	\rfoot{\footnotesize \bf Preprint of the manuscript accepted for publication in \emph{NAVIGATION}} \lfoot{\footnotesize \bf Copyright \copyright~2026 by Zachary L. Clements, Patrick B. Ellis, Iain \\ Goodridge, Matthew J. Murrian, Mark L. Psiaki, and Todd E. Humphreys}
	
	\else
	
	\thispagestyle{empty}
	\pagestyle{empty}
	
	\fi
	\pagestyle{plain}

\section{Introduction}

The combination of easily accessible low-cost GNSS spoofers and the emergence
of increasingly automated GNSS-reliant systems prompts a need for multi-layered
defenses against GNSS spoofing.  A GNSS spoofer emits an ensemble of false GNSS
signals intending that the victim receiver(s) accept them as authentic GNSS
signals, thereby inferring a false position fix and/or clock offset
\cite{psiaki2016gnssSpoofing, jafarnia2012vulReview}.  A successful spoofing
attack may lead to serious consequences.  

The academic community has long warned the public about the threat of GNSS
spoofing \cite{scott2003asa, t_humphreys_gcs08,
humphreys2012congressionalTestimony}.  Within the past decade, significant
progress has been made in GNSS spoofing detection and mitigation
\cite{psiaki2016gnssSpoofing, jafarnia2012vulReview, humphreysGNSShandbook,
psiakiNewBlueBookspoofing, rados2024recent}.  Reliable spoofing detection
techniques even exist for challenging environments such as dynamic platforms in
urban areas where strong multipath and in-band noise are common
\cite{wesson2018pincer,gross2018maximum, psiaki2014wrod,
ohanlon2012_IONrealTimeSpDet, gross2017gnss, b_ohanlon10_cgr}.  Consistency
checks between the estimated signal and onboard inertial sensors can provide
quick and reliable spoofing detection \cite{clements2022CpImuSpoofIonItm,
clements2023spoofingDetectionGpsWorld, tanil2018insMonitor, KujurINS2024}.
Monitoring the clock state can also be used to detect spoofing
\cite{jafarnia2013pvt, hwang2014receiver, khalajmehrabadi2018real}.
Cryptographic authentication techniques are currently being developed and
implemented to verify received signals \cite{humphreys2013ds,kerns2014nmaimp,
fernandez2023semi, MinaChimera2024, Andersonnavi.595, Andersonnavi.655}. 

Although the recent advances in GNSS spoofing detection have been inspiring,
many older GNSS receivers in current operation are unable to incorporate such
defenses, leaving them vulnerable to attacks.  For example, the civilian
maritime and airline industries are encountering GNSS jamming and spoofing at
an alarming rate \cite{c4ads2019aboveUsOnlyStars, gebrekidan2023nyt,
arraf2024npr, tangel2024wsj, FeluxAviation2024, opsgroup2024spoofing,
osechas2022impact}.  Anomalous positioning information broadcast by ships in
Automatic Identification System (AIS) messages, and airplanes in Automatic
Dependent Surveillance-Broadcast (ADS-B) messages, indicate recent widespread
jamming and spoofing.  These civilian aircraft and ships ensnared by GNSS
spoofing are likely unintended targets caught in the electronic warfare
crossfire near ongoing conflict zones.

GNSS spoofing attacks can be sorted into two categories, targeted spoofing and
broadcast spoofing. In targeted spoofing, an attacker transmits spoofing
signals for a specific (possibly moving) target it wishes to deceive.  This
type of attack involves the attacker tailoring a spoofing trajectory for its
specific target, causing a gradual pull-off from the victim's true trajectory,
and compensating for the relative motion between the spoofer and the target to
minimize target's probability of detection \cite{kerns2014unmanned}.  Targeted
spoofing is a sophisticated,  expensive, and difficult-to-detect attack that
requires the attacker to have the ability to precisely track the target and
craft spoofing signals in accordance with the target's motion, all in
real-time. Due to its complexity and narrow scope, this form of spoofing is the
least common.  Other GNSS receivers besides the targeted victim can also be
captured by these signals, but a non-targeted receiver can more easily detect
such spoofing.  Moreover, targeted spoofing may involve narrow beamforming,
making reception by non-target receivers unlikely.

Broadcast spoofing is less expensive, less complex, and wider in geographic
extent than targeted spoofing, and thus more common.  In broadcast spoofing, an
attacker transmits spoofing signals broadly with the intent to deceive all GNSS
receivers within a wide area.  Because broadcast spoofing is non-targeted,
victim GNSS receivers typically see a sudden jump in position and/or timing,
which is trivial to detect with basic spoofing detection checks.  Yet despite
being easy to detect, broadcast spoofing remains effective at denying GNSS
access to victims lacking proper defenses. When a GNSS receiver cannot
confidently differentiate between authentic and spoofing signals, it is
rendered useless---or worse: hazardously misleading.  The spoofers recently
affecting the aviation and maritime industries appear to be of the broadcast
type.


Given that many currently deployed GNSS receivers are unable to defend
themselves even against easy-to-detect broadcast spoofing, GNSS users need to
be warned of hazardous GNSS-challenged environments.  The proliferation of
LEO-based GNSS receivers provides the potential of unprecedented spectrum
awareness, enabling GNSS interference detection, classification, and
geolocation with worldwide coverage
\cite{lachapelle2021orbitalWardrivingIonGnss, murrian2021leo,
mckibben2023interference, clements2023PlansDirectGeo,
clements2023pinpointingInterferenceInsideGNSS, ChewRFI2023}.  Existing and
proposed LEO constellations provide worldwide coverage with frequent revisit
rates, allowing for an always-updating operating picture, a noted shortfall in
current capabilities \cite{navwar2024NSSA}.  Several commercial enterprises
have seized the opportunity to deploy constellations of LEO satellites to
provide spectrum monitoring and emitter geolocation as a service (e.g., Spire
Global and Hawkeye360).

With multiple time-synchronized receivers, geolocation of emitters producing
arbitrary wideband signals is possible and has been extensively studied
\cite{sidi14_dpf, musicki10_meg, ho1997geolocation,
clements2023PlansDirectGeo, clements2023pinpointingInterferenceInsideGNSS}.
Multiple time-synchronized receivers can exploit time- and
frequency-difference-of-arrival (T/FDOA) measurements to estimate the emitter
location.  The authors of the current paper were able to geolocate over 30 GNSS
interference sources across the Near East from a dual-satellite
time-synchronized capture \cite{clements2023PlansDirectGeo,
clements2023pinpointingInterferenceInsideGNSS}.  However, planning simultaneous
multi-satellite captures to enable T/FDOA-based geolocation can be difficult to
coordinate and expensive, whereas single-satellite collects are straightforward
and less costly.  Accordingly, this paper focuses on single-satellite
geolocation.


Accurate single-satellite geolocation of emitters with arbitrary waveforms is
impossible in general: if the signal’s carrier cannot be tracked, only coarse
received-signal-strength techniques can be applied.  But if a signal's carrier
can be tracked, or Doppler can be otherwise measured, then accurate
single-satellite-based emitter geolocation is possible from Doppler
measurements alone, provided that the emitter's carrier frequency is
quasi-constant \cite{ellis2020use, murrian2021leo, ellis2018performance,
ellis2020errors}.  But if a transmitter introduces any significant level of
complexity to the carrier-phase behavior, such as frequency modulation or clock
dithering, the accuracy of Doppler-based single-satellite techniques degrades.  

GNSS spoofers must be treated specially, as they do not transmit at a constant
carrier frequency: they add an unknown time-varying frequency component to each
spoofing signal, imitating the range-rate between the corresponding spoofed
GNSS satellite and the counterfeit spoofed location \cite{kerns2014unmanned}.
A key contribution of the current paper is a technique that removes the unknown
time-varying frequency component added by GNSS spoofers so that a range-rate
time history can be extracted for geolocation.  \cite{chen2023gnss} also
presented a single-receiver spoofer geolocation technique based on counterfeit
clock observables. However, \cite{chen2023gnss} only considers the spoofed
pseudorange measurements and depends on a stationary receiver initialization
period, which is not possible in LEO.  

The key observation behind this paper's technique is that each spoofed
navigation signal will share a common frequency shift due to the range-rate
between the LEO receiver and the terrestrial spoofer.  If a GNSS receiver
processes enough spoofing signals to form a navigation solution, then the
receiver's internal estimator will naturally lump the common frequency shift of
each signal from the shared range-rate into the receiver clock drift (clock
offset rate) estimate.  Therefore, the time history of the spoofed receiver
clock drift can be exploited for geolocation because the range-rate between the
LEO receiver and the terrestrial spoofer is embedded in this measurement.


This paper makes four primary contributions. First, it presents a
single-satellite, single-pass GNSS spoofer geolocation technique that extracts
a range-rate between a LEO-based receiver and a terrestrial broadcast spoofer
from captured raw samples.  Second, it offers an experimental demonstration of
the technique with a truth solution.  Third, it derives an analytic expression
for how transmitter clock instability degrades the single-satellite geolocation
solution. Fourth, it investigates the geolocation positioning errors as a
function of worst-case spoofed clock behavior.

Preliminary conference versions of this paper were published in
\cite{clements2022spoofergeo, clements2024DemoSpooferGeo}.  The current
version significantly extends these with contributions three and four mentioned
above.

\section{Signal Models}
\label{sec:signals}
\subsection{GNSS Spoofing Signals}
The goal of a broadcast GNSS spoofer is to deceive the victim receiver(s) into
inferring a false position, velocity, and timing (PVT) solution, denoted
$\vb{\tilde{x}} = [ \:\vb{r}_{\tilde{\text{r}}}^\tr, \; \delta
t_{\tilde{\text{r}}}, \; \vb{v}_{\tilde{\text{r}}}^\tr, \; \delta
\dot{t}_{\tilde{\text{r}}} \:]^\tr$, where $\vb{r}_{\tilde{\text{r}}}$ is the
spoofed position in Earth-centered-Earth-fixed (ECEF) coordinates, $\delta
t_{\tilde{\text{r}}}$ is the spoofed clock bias increment,
$\vb{v}_{\tilde{\text{r}}}$ is the spoofed velocity, and $\delta
\dot{t}_{\tilde{\text{r}}}$ is the spoofed clock drift increment.  To achieve a
successful attack, the spoofer must generate an ensemble of self-consistent
signals.  To this end, the attacker must (1) select a counterfeit PVT solution
for the victim to infer, (2) select an ensemble of GNSS satellites to spoof,
and (3) for each spoofed navigation satellite, generate a signal with a
corresponding navigation message, code phase time history, and carrier phase
time history consistent with (1) and (2).


A general baseband signal model for broadcast spoofing signals is now
presented.  The ensemble of spoofing signals transmitted by the spoofer,
denoted 
\begin{align}
	x(t) = \sum_{n=1}^{N}s_n(t)
\end{align}
contains $N$ spoofing signals, where the $n$th spoofing signal is denoted $s_n(t)$ for $n = 1,2, ..., N$.
The $n$th  spoofing baseband signal takes the form 
\begin{align}
	s_n(t) = A_nD_n\left[t - \tau_n(t)\right]C_n\left[t - \tau_n(t)\right]\exp\left[j2\pi\theta_n(t)\right]
\end{align}
where $A_n$ is the carrier amplitude, $D_n(t)$ is the data bit stream, $C_n(t)$
is the  spreading code, $\tau_n(t)$ is the code phase, and $\theta_n(t)$ is the
negative beat carrier phase \cite{psiaki2016gnssSpoofing}.  The Doppler of the
$n$th spoofing signal is related to $\theta_n(t)$ by
\begin{align}
	\tilde{f}_n(t) = \frac{d}{dt}\theta_n(t)
\end{align}
The spoofer adds a unique Doppler component to each spoofing signal that mimics
the combined Doppler of the following components: (1) the range-rate between
the spoofed satellite and spoofed position, (2) the spoofed receiver clock
drift, and (3) the spoofed satellite clock drift.  Additionally, the spoofed
code phase and carrier phase time histories must be mutually consistent to
avoid code-carrier divergence.  Accordingly, the Doppler of the $n$th
transmitted spoofing signal may be modeled as
\begin{align}
	\tilde{f}_{n}(t) = &- \frac{1}{\lambda} \hat{\vb{r}}_{\tilde{\text{r}}n}^\tr(t) \left(\vb{v}_{\tilde{\text{r}}}(t) - \vb{v}_{\tilde{\text{s}}n}(t)\right) - \frac{c}{\lambda} \left(\delta \dot{t}_{\tilde{\text{r}}}(t) -  \delta \dot{t}_{\tilde{\text{s}}n}(t)\right)
	\label{eq:spooferDopplerTX}
\end{align}
where $\lambda$ is the carrier wavelength, $c$ is the speed of light,
$\hat{\vb{r}}_{\tilde{\text{r}}n}$ is the unit vector pointing from the $n$th
spoofed navigation satellite to the spoofed position, both in ECEF coordinates,
$\vb{v}_{\tilde{\text{r}}}$ is the spoofed receiver velocity,
$\vb{v}_{\tilde{\text{s}}n}$ is the $n$th spoofed navigation satellite
velocity, and $\delta \dot{t}_{\tilde{\text{s}}n}$ is the spoofed clock drift
of the $n$th navigation satellite.  One can immediately appreciate that the
Doppler frequency is different for each spoofing signal.  Had this been a
targeted spoofer, there would be an additional Doppler term in
(\ref{eq:spooferDopplerTX}) that compensates for the relative motion between
the victim and spoofer, but in the case of broadcast spoofing, this term is
zero.

\subsection{Received Doppler Model}

First consider a scenario in which a moving receiver captures a transmitted
signal having a constant carrier frequency.  The received Doppler
$f_\text{D}(t)$ at the moving receiver can be modeled as 
\begin{align}
	f_\text{D}(t) = & - \frac{1}{\lambda} \hat{\vb{r}}^\tr(t) \left(\vb{v}_{\text{r}}(t) - \vb{v}_{\text{t}}(t)\right) - \frac{c}{\lambda} \left(\delta \dot{t}_\text{r}(t) - \delta \dot{t}_\text{t}(t) \right)
	\label{eq:fD}
\end{align}
where $\hat{\vb{r}}$ is the unit vector pointing from the transmitter to the
receiver, $\vb{v}_\text{r}$ is the velocity of the receiver, $\vb{v}_\text{t}$
is the velocity of the transmitter, $\delta \dot{t}_\text{r}$ is the clock
drift of the receiver, and $\delta \dot{t}_\text{t}$ is the clock drift of the
transmitter. Note that this is a simplified Doppler model that neglects
higher-order terms.  \cite{psiaki2021navigation} presented a complete Doppler
model.  For the purposes of this paper, the simplified model is adequate, as
will be confirmed by the experimental results.

\begin{figure}[t]
	\centering
	\includegraphics[width=.8\linewidth]{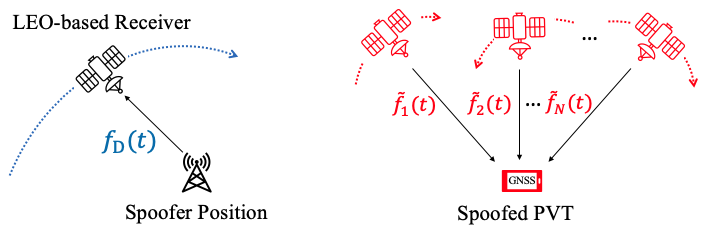}
	\caption{Shown here are the Doppler components in single-satellite spoofer geolocation.  The Doppler  components corresponding to (\ref{eq:fD}) are shown on the left. The Doppler components for each spoofing signal corresponding to (\ref{eq:spooferDopplerTX}) are shown in red to the right.  }
	\label{fig:geometry}
\end{figure}

Now consider a scenario in which a moving receiver captures an ensemble of
transmitted spoofing signals from a stationary terrestrial spoofer
($\vb{v}_{\text{t}}(t)$ = 0), as shown in Fig.~\ref{fig:geometry}.
\cite{clements2022spoofergeo} provided an analysis of how spoofer motion
affects the geolocation solution.  But would-be spoofers are typically
stationary; otherwise, they face the additional difficulty of compensating for
their motion to avoid producing easily-detectable false signals.  Therefore, a
stationary spoofer will be assumed for the rest of this paper.

Each observed signal at the receiver will contain a common Doppler shift
$f_\text{D}$ due to the the relative motion between the transmitter (spoofer)
and the receiver.  Each observed signal will also manifest a common frequency
shift due to the clock drift of the transmitter and the clock drift of the
receiver.  Dropping time indices for clarity, the observed Doppler of the $n$th
spoofing signal at the moving receiver, $f_n$, may be written as  
\begin{align}
	f_{n} = &f_\text{D} + \tilde{f}_n \nonumber \\
	=  &- \frac{1}{\lambda} \hat{\vb{r}}^\tr \vb{v}_{\text{r}} - \frac{c}{\lambda} \left(\delta \dot{t}_\text{r} - \delta \dot{t}_\text{t} \right) \nonumber \\
	   &- \frac{1}{\lambda} \hat{\vb{r}}_{\tilde{\text{r}}n}^\tr \left(\vb{v}_{\tilde{\text{r}}} - \vb{v}_{\tilde{\text{s}}n}\right) - \frac{c}{\lambda} \left(\delta \dot{t}_{\tilde{\text{r}}} -  \delta \dot{t}_{\tilde{\text{s}}n}\right)
	\label{eq:observedDoppler}
\end{align}

What makes single-satellite GNSS spoofer geolocation difficult is the
$\tilde{f}_n$ term: it is typically unknown, time-varying, and different for
each spoofing signal.  In the case of the matched-code jammer discovered by
\cite{murrian2021leo}, $\tilde{f}_n = 0$.  One may suppose that the operator's
intent in that case was not to deceive victim receivers into inferring false
locations like a spoofer.  When $\tilde{f}_n = 0$, the observed Doppler can be
modeled as the range-rate between transmitter and receiver, with a constant
measurement bias over the capture to account for the clock drift of the
transmitter.  Contrariwise, naive geolocation with the observed Doppler modeled
as in (\ref{eq:observedDoppler}) yields final position estimates that are
biased because the spoofing signals contain the unmodeled $\tilde{f}_n(t)$
term.  In the following section, a technique is presented that removes
$\tilde{f}_n(t)$ and extracts $\hat{\vb{r}}^\tr(t) \vb{v}_{\text{r}}(t)$, the
range-rate time history between transmitter and receiver, which can be
exploited for geolocation.

\section{Conceptual Overview of Broadcast GNSS Spoofer Geolocation}
\label{sec:conceptual} This section presents an overview of the technique for
spoofer geolocation originally presented by \cite{clements2022spoofergeo,
clements2024DemoSpooferGeo}.  The common Doppler components across all spoofing
signals from (\ref{eq:observedDoppler}) are indicated below:
\begin{align}
  f_{n} = &\underbrace{ - \frac{1}{\lambda} \hat{\vb{r}}^\tr \vb{v}_{\text{r}} - \frac{c}{\lambda} \left(\delta \dot{t}_\text{r} - \delta \dot{t}_\text{t} \right)}_{\text{common}}  \nonumber \\
          &- \frac{1}{\lambda} \hat{\vb{r}}_{\tilde{\text{r}}n}^\tr \left(\vb{v}_{\tilde{\text{r}}} - \vb{v}_{\tilde{\text{s}}n}\right) - \frac{c}{\lambda} (\underbrace{\delta \dot{t}_{\tilde{\text{r}}}}_\text{common} -  \delta \dot{t}_{\tilde{\text{s}}n}) 
           \label{eq:obsDoppler1}
\end{align}
All common Doppler terms can be lumped into a single term
\begin{align}
  \gamma(t) = \frac{1}{c} \hat{\vb{r}}^\tr (t) \vb{v}_{\text{r}}(t) + \delta \dot{t}_\text{r}(t) - \delta \dot{t}_\text{t}(t)  + \delta \dot{t}_{\tilde{\text{r}}} (t)
  \label{eq:gamma}
\end{align}
so that (\ref{eq:observedDoppler}) may be written
\begin{align}
  f_{n}= - \frac{1}{\lambda} \hat{\vb{r}}_{\tilde{\text{r}}n}^\tr \left(\vb{v}_{\tilde{\text{r}}} - \vb{v}_{\tilde{\text{s}}n}\right) - \frac{c}{\lambda} \left(\gamma - \delta \dot{t}_{\tilde{\text{s}}n}\right) 
  \label{eq:obsDoppler3}
\end{align}

Upon processing an ensemble of spoofing signals, a GNSS receiver's PVT
estimator produces, at each navigation epoch, the state estimate
\begin{equation}
  \label{eq:stateEstimate}
\hat{\vb{x}}(t) = [\:\hat{\vb{r}}^\tr_{\tilde{\text{r}}}(t), \; \hat{\xi}(t),
\; \hat{\vb{v}}^\tr_{\tilde{\text{r}}}(t), \; \hat{\gamma}(t) \:]^\tr  
\end{equation}
which is composed of the estimated spoofed position, the estimated receiver
clock bias $\hat{\xi}(t)$, the estimated spoofed velocity, and the estimated
receiver clock drift $\hat{\gamma}(t)$ \cite{gunther2014survey}.
\cite{clements2024DemoSpooferGeo, odijkGNSShandbook} provide a brief review of
PVT estimation from pseudorange and Doppler measurements.  Note that the
estimated receiver clock bias $\hat{\xi}(t)$ will include $\delta
t_{\tilde{\text{r}}}$ as a component but will not in general equal $\delta
t_{\tilde{\text{r}}}$.

The estimated clock drift $\hat{\gamma}(t)$, on the other hand, will track
$\gamma(t)$ closely provided that the PVT estimator is configured with a clock
model whose process noise intensity is sufficient to accommodate the variations
in $\gamma(t)$ due to spoofing.  Expressed in s/s, $\hat{\gamma}(t)$ contains
all common Doppler terms, since the PVT estimator attributes common-mode
frequency deviations across received signals to the receiver's clock drift.
Importantly, $\hat{\gamma}(t)$ is unaffected by the unknown non-common Doppler
components from $\tilde{f}_n(t)$ for all $n \in \{ 1, 2, ..., N \}$.

The time history $\hat{\gamma}(t)$ is the key to spoofer geolocation because it
depends strongly on the range-rate between the LEO-based receiver and the
terrestrial spoofer.  In particular, information about the transmitter's
location is embedded in $\hat{\vb{r}}^\tr (t) \vb{v}_{\text{r}}(t)$, which for
a LEO-based receiver is typically the dominant component in $\gamma(t)$.  A
nonlinear least-squares estimator based on $\hat{\gamma}(t)$ is developed in
the next section to estimate the spoofer's position.

The other three terms in $\gamma(t)$, namely $\delta \dot{t}_\text{r}(t)$,
$\delta \dot{t}_\text{t}(t)$, and $\delta \dot{t}_{\tilde{\text{r}}}(t)$, are
nuisance terms that potentially degrade geolocation accuracy.  Fortunately,
their contributions are typically minor or can be estimated.  Consider $\delta
\dot{t}_\text{r}(t)$.  If the satellite's GNSS receiver and the radio frequency
(RF) front-end capturing spoofing signals are driven by the same oscillator,
then $\delta \dot{t}_\text{r}(t)$ is automatically estimated by the onboard
GNSS receiver, provided it is not significantly affected by the spoofing, and
and thus $\delta \dot{t}_\text{r}(t)$ can be compensated.

It is worth mentioning that one of the core assumptions in any geolocation
system is that the capture platform has knowledge of its PVT; otherwise,
geolocation is impossible.  In the scenario assumed in this paper, the
LEO-based receiver has access to its PVT from an onboard GNSS receiver that is
robust to terrestrial interference.  Despite the presence of spoofing signals,
code- and carrier-tracking of the authentic GNSS signals is maintained due to
sufficient separation of the false and authentic signals in the code-Doppler
space, as achieved by \cite{murrian2021leo}.  Furthermore, robustness is
achieved if a zenith-facing antenna feeds the onboard GNSS receiver's RF
front-end, as the gain directed towards Earth will be strongly attenuated.  PVT
can be trivially maintained by a multi-constellation receiver when only
single-constellation spoofing signals are present.  In the event that all GNSS
signals are unavailable due to terrestrial interference, knowledge of the
receiver's position and velocity can be maintained by using orbital propagation
models such as simplified general perturbations 4 (SGP4).  Over short periods,
the orbit is stable enough so that the receiver will be able to maintain
sufficient PVT accuracy from the onset of GNSS denial.  

The terms $\delta\dot{t}_\text{t}(t)$ and $\delta
\dot{t}_{\tilde{\text{r}}}(t)$ originate from the spoofer.  Specifically,
$\delta\dot{t}_\text{t}(t)$ originates from the spoofer's hardware, while
$\delta \dot{t}_{\tilde{\text{r}}} (t)$ originates from its spoofer's software.
The former arises due to the clock drift in the spoofer. It can often be
accurately modeled as constant over short (e.g., 60-second) capture intervals
and estimated as part of the geolocation process \cite{murrian2021leo}.  The
spoofed clock drift $\delta \dot{t}_{\tilde{\text{r}}}(t)$ arises from the
spoofer's attack configuration, and will manifest at the victim as an increment
to the victim's clock drift.  It can be troubling for geolocation, but a
potential attacker would typically opt to keep $\delta
\dot{t}_{\tilde{\text{r}}}(t)$ near constant, because if
$\delta\dot{t}_{\tilde{\text{r}}}(t)$ grows too rapidly to be explained by the
expected variation in clock drift for the receiver’s oscillator type, the
victim receiver could flag the anomaly and thereby detect the spoofing attack. 

This constraint can be generalized to the sum $\delta\dot{t}_\text{t}(t)  +
\delta \dot{t}_{\tilde{\text{r}}} (t)$ and summarized as follows: if the
spoofer allows extraordinary frequency instability in its own oscillator so
that $\delta\dot{t}_\text{t}(t)$ changes too rapidly, or if it attempts to
induce a quickly-varying spoofed clock drift so that
$\delta\dot{t}_{\tilde{\text{r}}} (t)$ changes too rapidly, geolocation
accuracy is degraded but, on the other-hand, the spoofing attack becomes
trivially detectable.

Additionally, for a targeted spoofing attack in which the spoofer attempts to
compensate for true spoofer-to-victim line-of-sight velocity, $\gamma(t)$ could
contain an extra time-varying term.  If this term were to vary rapidly with
time, it would cause trouble for this paper's technique.  Relatedly, if the
targeted victim's position and velocity were somehow accurately known to the
LEO-based receiver, this paper's technique could produce accurate results
provided that the estimator presented in the next section were updated to
account for the known victim motion.  Finally, if the targeted victim receiver
is stationary, this paper's technique can be applied without modification.

Section \ref{sec:error_analysis} explores the consequences for geolocation of
cases where $\delta\dot{t}_\text{t}(t)$ departs from a constant model.  It also
presents an analysis of how aggressively an attacker can ramp $\delta
\dot{t}_{\tilde{\text{r}}}(t)$ without being detected by an optimal spoofing
detection strategy that monitors the receiver clock drift, and an analysis of
how the rate of change in $\delta\dot{t}_\text{t}(t) + \delta
\dot{t}_{\tilde{\text{r}}} (t)$ translates to geolocation error.  

\section{Spoofer Geolocation with $\gamma$}
\label{sec:estimator}
This section presents the measurement model, derives the measurement noise
covariance matrix, and presents the nonlinear least-squares estimator for
single-satellite spoofer geolocation.

\subsection{Measurement Model}
\label{ssec:measurement_model}

When a GNSS receiver processes spoofing signals, it first generates spoofed
GNSS observables.  These GNSS observables are beset with errors, modeled as
zero-mean additive white Gaussian noise (AWGN), arising from thermal noise,
local electromagnetic interference, atmospheric and relativistic effects,
ephemeris errors, and other minor effects.  At every navigation epoch, the
noisy spoofed GNSS observables are fed to the receiver's PVT estimator to
produce an optimal estimate of the spoofed PVT solution, including
$\hat{\gamma}(t)$.

Let ${\gamma}[i] = {\gamma}(i\Delta t)$ and $\hat{\gamma}[i] =
\hat{\gamma}(i\Delta t)$, where $\Delta t$ is the constant PVT solution
interval and $i \in \mathcal{I} = \{ 1,2, ..., I \}$ is the solution index
within a given data capture interval.  Let $z[i]$ denote the $i$th measurement
to be used for spoofer geolocation, modeled as
\begin{align}
  z[i] = c\hat{\gamma}[i] = c\gamma[i] + w_\text{a}[i], \quad i \in \mathcal{I}
  \label{eq:z}
\end{align}
The velocity-equivalent estimation error $w_\text{a}[i]$, which has units of
m/s, is a discrete-time noise process with $\E{w_\text{a}[i]} = 0$ and
$\E{w_\text{a}[i]w_\text{a}[j]} = \sigma^2_\text{a}\delta_{ij}, ~ \text{for all
} i,j \in \mathcal{I}$.  Section \ref{subsec:correlatederrors} will justify
this model's assumption that $w_\text{a}[i]$ is white (uncorrelated in time)
for a sufficiently large $\Delta t$ that is larger than the settling time of
its phase lock loop (PLL) or frequency lock loop (FLL), and the settling time
of any Kalman filter used for obtaining the spoofed fix.

As stated before, $\delta\dot{t}_\text{r}$ is assumed to be known and fully
compensated; accordingly, it will be neglected hereafter.  Additionally,
$\delta \dot{t}_{\tilde{\text{r}}}$ is part of the spoofer's attack
configuration and, for now, will be modeled as constant due to the constraints
mentioned in the prior section.

A more comprehensive model is considered for $\delta\dot{t}_\text{t}(t)$.  Let
$\delta\dot{t}_\text{t}[i] = \delta\dot{t}_\text{t}(i\Delta t), i \in
\mathcal{I}$.  Over a capture interval, $\delta\dot{t}_\text{t}[i]$ is modeled
as
\begin{align}
  c\delta\dot{t}_\text{t}[i] = c\delta\dot{t}_\text{t}[0] + b[i], \quad i \in \mathcal{I}
\end{align}
where $\delta\dot{t}_\text{t}[0]$ represents the spoofer oscillator's constant
frequency bias and $b[i]$ is a Gaussian random walk process expressed as
\begin{align}
	b[i] = \sum_{k=1}^i v[k], \quad i \in \mathcal{I}
\end{align}
where $v[k]$ is a discrete-time Gaussian random process with $\E{v(k)} = 0$,
$\E{v[k]v[j]} = \sigma^2_v \delta_{kj}$ and
$\E{w_\text{a}[k]v[j]}~=~0 ~ \text{for all } k,j \in \mathcal{I}$, and
$b[0] = 0$.  Using the model in \cite[Chap.  8]{brown2012introKf}, $\sigma_v^2$
can be characterized as
\begin{align}
  \sigma^2_v = 2 \pi^2 h_{-2} \Delta t c^2 
  \label{eq:sig_v}
\end{align}
where $h_{-2}$ is the first parameter of the standard clock model based on the
fractional frequency error power spectrum \cite{murrian2021leo}.  Scaling by
$c^2$ converts to units of $\text{(m/s)}^2$.  

Note that $\delta\dot{t}_\text{t}[0]$ and $\delta \dot{t}_{\tilde{\text{r}}}$
can be combined into a single measurement bias $b_0$ that is constant across
the capture interval.  Furthermore, the AWGN and Gaussian random walk can also
be combined into a single noise term $w[i]$. Thus we have
\begin{align}
	b_0 &= -c\delta\dot{t}_\text{t}[0] + c\delta \dot{t}_{\tilde{\text{r}}} \\
	w[i] &= w_\text{a}[i] + b[i], \quad i \in \mathcal{I}
\end{align}
Given all of this, (\ref{eq:z}) is rewritten so that the final measurement
model takes the form 
\begin{align}
	z[i] = \hat{\vb{r}}^\tr_i \vb{v}_{\text{r},i} + b_0 + w[i], \quad i \in \mathcal{I}
	\label{eq:z_final}
\end{align}

The associated measurement covariance matrix $R$ for the process $w[i]$ is now
derived.  Clearly, $w[i]$ is zero-mean, but because it contains a Gaussian
random walk term, it is correlated over time.  The $[i,j]$th element of its
measurement covariance matrix is 
\begin{align}
	R[i,j] &= \E{w[i]w[j]} \nonumber \\
	&= \E{\left(w_\text{a}[i] + \sum_{k=1}^{i}v[k]\right) \left( w_\text{a}[j] + \sum_{l=1}^{j}v[l]\right)} \nonumber \\
	&= \E{w_\text{a}[i]w_\text{a}[j]} + \E{\left(\sum_{k=1}^{i}v[k]\right)\left(\sum_{l=1}^{j}v[l]\right)} \nonumber \\
	&= \E{w_\text{a}[i]w_\text{a}[j]} + \sum_{k=1}^{i}\sum_{l=1}^{j}\E{v[k]v[l]} \nonumber \\
	&= \sigma^2_\text{a}\delta_{ij} + \sigma_v^2 \;  \text{min}\{i, j\}
\end{align}
From this result, the measurement covariance matrix containing the AWGN and
Gaussian random walk can be  written as 
\begin{align} 
	R &= R_\text{a} +R_\text{b}  \quad \text{where} \quad 
	\label{eq:R}
	R_\text{a} = \sigma^2_\text{a} \; \mathds{I}_{I\times I} \quad \text{and} \quad
	R_\text{b} = \sigma^2_v \; M_{I\times I}
\end{align}
where $\mathds{I}_{I\times I}$ is the identity matrix and $M$ is an $I\times I$
matrix with $M[i,j] = \text{min}\{i, j\}, ~ i \in \mathcal{I}$.  Note that this
covariance matrix is a general result that can be applied to any
range-rate-based positioning technique where the transmitter clock state is
unknown.

\subsection{Effects of Estimated $\gamma$}
\label{subsec:correlatederrors}

One might question the choice to model the estimation error process
$w_\text{a}[i] = c(\hat{\gamma}[i] - \gamma[i])$ as white, since
$\hat{\gamma}[i]$ is the product of a state estimator and it is well known that
state estimation errors are correlated in time.  At epoch $i$, let $\tilde{\vb
x}[i]$ denote the sequential PVT estimator's full state estimation error,
$W[i]$ its feedback gain, $F[i]$ its state transition matrix, and $P[i]$ its
state covariance.  The covariance between sequential state errors is given by
\cite[Chap. 5]{y_barshalom01_tan} 
\begin{align}
 \E{\tilde{\vb x}[i+1]\tilde{\vb x}^\tr[i]} = \left(\mathds{I} - W[i+1]H[i+1]\right)F[i]P[i]
\end{align}
The correlation between $w_\text{a}[i+1]$ and $w_\text{a}[i]$ for $i \in
\mathcal{I}$ can be determined by analysis of this equation since
$w_\text{a}[i]$ is an element of $\vbt{x}[i]$.

Consider a scenario where the spoofer induces a static location with a typical
GPS satellite geometry.  The state estimated by an affected receiver consists
of the position, clock bias, and clock drift, as in (\ref{eq:stateEstimate}).
Assume the receiver's PVT estimator applies a dynamics model consistent with a
static position and the clock process noise model from
\cite{brown2012introKf}.  Furthermore, assume that measurement errors are
independent, zero-mean, and Gaussian with standard deviations of 1~m and
0.5~m/s respectively for the spoofed pseudorange and Doppler measurements.

\begin{figure}[t]
  \centering
  \includegraphics[width=.98\linewidth]{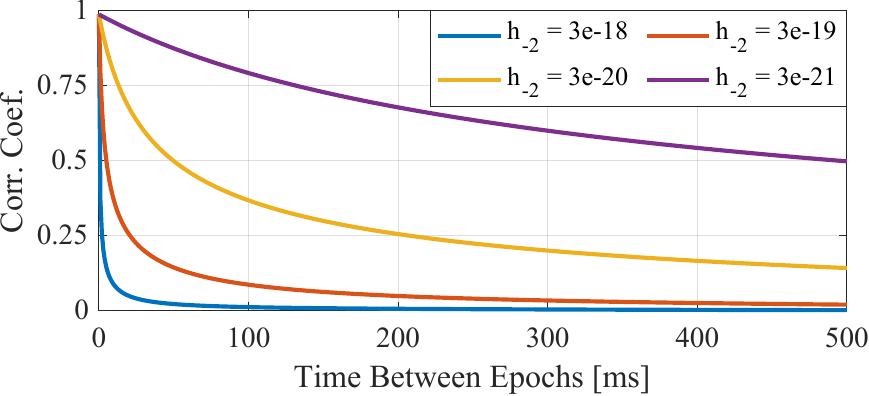}
  \caption{The Pearson correlation coefficient between sequential estimation
    errors $w_\text{a}[i]$ as a function of time between estimation epochs for
    various values of $h_{-2}$.  As the receiver's modeled process noise
    intensity increases, the time correlation between between estimation errors
    decreases.}
  \label{fig:correlation}
\end{figure}

A key tuning parameter in this model is the process noise of the receiver clock
drift, which is governed by the $h_{-2}$ coefficient, as in (\ref{eq:sig_v}).
Fig. \ref{fig:correlation} shows the Pearson correlation coefficient for
$w_\text{a}[i]$ between subsequent navigation epochs over various values of
modeled $h_{-2}$ as a function of the time between epochs.  As the process
noise and time between epochs is increased, the time correlation of sequential
estimation errors is reduced.  This type of analysis can be performed to help
determine the measurement interval length beyond which errors in the sequential
estimates $\hat{\gamma}[i]$ can be accurately approximated as AWGN.  For
example, Fig.~\ref{fig:correlation} indicates that, for $h_{-2} \ge 3\times
10^{-19}$, measurements spaced by 100 ms or more may be treated as independent.

If $h_{-2}$ were increased even further, the navigation filter becomes a
sequence of point solutions and, in effect, the white noise-model of
$w_\text{a}[i]$ is undoubtedly correct.  The selection of $h_{-2}$ becomes a
tuning parameter for the system designer.  This analysis involving nominal
$h_{-2}$ values becomes relevant because currently deployed LEO-based GNSS
receivers can perform this technique and may not have the flexibility to change
their own process noise.

\subsection{Range-rate Nonlinear Least-Squares Estimator}
\label{subsec:nonlinear_LS}
Now that the measurements and the measurement covariance have been defined, a
batch nonlinear least-squares estimator may be developed to solve for the state
$\vb x$
\begin{align}
  \vb{x} = 
  \begin{bmatrix}
    \vb{r}_\text{t}  \\
    b_0
  \end{bmatrix}
\end{align}
where $\vb{r}_\text{t}$ is the transmitter's ECEF position and $b_0$ is the
unknown measurement bias. Let $\vb z$ represent the $I \times 1$ stacked
measurement vector.  The standard weighted nonlinear least-squares cost
function is  
\begin{align}
  J(\vb x) = \frac{1}{2} \left[ \vb z - \vb{h(x)} \right]^\tr R^{-1}\left[ \vb z - \vb{h(x)} \right]
\end{align}
where $\vb{h}(\vb x)$ is the nonlinear measurement model function.  The optimal
estimate of $\vb x$ minimizes the cost $J$.  

The linearized measurement model $H$ is an $I \times 4$ matrix that takes the
form
\begin{align}
	H = 
	\begin{bmatrix}
		\frac{d h_1}{d \vb{r}_\text{t}}  \quad &1 \\
		\vdots \quad &\vdots \\
		\frac{d h_I}{d \vb{r}_\text{t}} \quad &1
	\end{bmatrix}
\end{align}
where 
\begin{align}
	\frac{d h_i(\vb x)}{d \vb{r}_\text{t}} =  \vb{v}_{\text{r},i}^\tr \frac{\left( \hat{\vb{r}}_{i} \hat{\vb{r}}_{i}^\tr - \mathds{I}_{3 \times 3}\right)}{\rho_i}
\end{align}
is the $1 \times 3$ Jacobian  of the $i$th range-rate measurement.  The range
between the receiver and the transmitter at the $i$th measurement is denoted
$\rho_i$.  This measurement model Jacobian is equivalent to columns 1, 2, 3,
and 8 of the Jacobian presented by \cite{psiaki2021navigation}, up to a scale
factor.

Enforcing an altitude constraint significantly improves the problem's
observability. This can be incorporated as an additional pseudo-measurement of
the transmitter's altitude with respect to the WGS-84 ellipsoid, modeled as
\begin{align} z[I+1] = h_\text{alt}(\vb{x}) + w_\text{alt} \end{align} where
the measurement error $w_\text{alt} ~\sim
\mathcal{N}(0,\,\sigma_\text{alt}^{2})$ is assumed to be independent of those
for $z[i], i \in \mathcal{I}$. The measurement's 1~$\times$~4 Jacobian is
\begin{align}
	H_\text{alt} =  
	\begin{bmatrix}
		\text{cos}(\phi_\text{lat})\text{cos}(\lambda_\text{lon}) , \; \text{cos}(\phi_\text{lat}) \text{sin}(\lambda_\text{lon}) , \;  \text{sin}(\phi_\text{lat}) , \; 0
	\end{bmatrix}
\end{align} 
where $\phi_\text{lat}$ and $ \lambda_\text{lon}$ are the latitude and
longitude of $\vb{r}_\text{t}$, respectively. The measurement vector $\vb{z}$,
vector-valued function $\vb{h}(\vb{x})$, Jacobian $H$, and error covariance $R$
are all augmented appropriately to include the altitude pseudo-measurement.

Finally, the estimation error's Cram\'er-Rao lower bound (CRLB) can be
approximated as 
\begin{align}
	P_{\vb{x}\vb{x}} = \left( H^\tr R^{-1}H\right)^{-1}
	\label{eq:Pxx}
\end{align}

\section{Experimental Results}
\label{sec:experimental_results}
The single-satellite geolocation technique described above was verified in a
joint demonstration between the University of Texas Radionavigation Laboratory
(UT RNL) and Spire Global.  In this experiment, an ensemble of self-consistent
spoofing signals was transmitted from a ground station while an overhead LEO
satellite performed a raw signal capture.  This section details the
experimental setup and results.  Preliminary results were presented by
\cite{clements2024DemoSpooferGeo}, which contains a comprehensive description
of the special adaptations made to deal with the spoofer's non-GNSS carrier
frequency. 

\subsection{Experimental Design}

The UT RNL provided a baseband binary file containing an ensemble of GNSS
spoofing signals to be transmitted, a filtered and downsampled version of the
``clean static"  recording in the TEXBAT dataset
\cite{humphreys2012_TEST_Battery}.  The original recording was a high-quality
16-bit 25 Msps (complex) recording of authentic GNSS signals centered at GPS L1
from a stationary antenna on top of the former Aerospace Engineering building
at UT Austin. The front-end in the original recording was driven by a 10-MHz
oven-controlled crystal oscillator (OCXO).  Lowpass filtering and downsampling
of the original file was required to ensure the transmitted signal was
contained within Spire's available bandwidth.  Additionally, onboard the
satellite, the S-band capture device and onboard GNSS receiver were driven by
the same oscillator, allowing precise time-tagging and compensation.

\begin{figure*}[t]
	\centering
	\begin{minipage}[b]{0.32\textwidth}
		\centering
		\includegraphics[width=\linewidth]{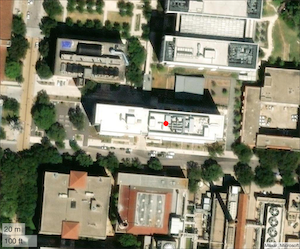}
	\end{minipage}
	\begin{minipage}[b]{0.32\textwidth}
		\centering
		\includegraphics[width=\linewidth]{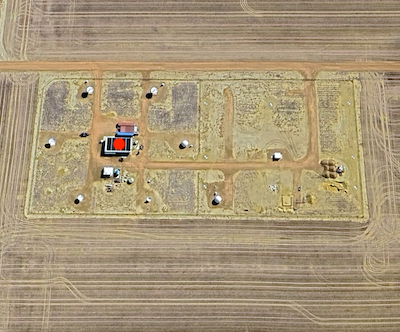}
	\end{minipage}
	\begin{minipage}[b]{0.32\textwidth}
		\centering
		\includegraphics[width=\linewidth]{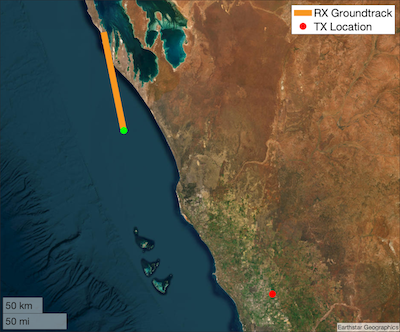}
	\end{minipage}
	\caption{Left: The spoofed location atop
	the former Aerospace Engineering building in Austin, Texas.  Center: The actual spoofer location, a Spire Global ground station located in Perth, Australia.  Right: The ground track of the Spire Global LEO satellite during the 20-second signal capture.}
	\label{fig:experimentalSetup}
\end{figure*}

The spoofing file was transmitted from a ground station located in Perth,
Australia.  The transmitter was driven by a temperature-controlled crystal
oscillator (TCXO).  The transmitted spoofing signals were centered at S-band to
avoid interfering with the GNSS bands.  While the ground station was
transmitting the spoofing file, an overhead LEO satellite performed a raw
signal capture over 20 seconds, centered at the S-band carrier and sampled at 5
Msps (complex). In practice, all processing would be done by an onboard
receiver. The duration of the raw capture should be as long as a frame in the
spoofed navigation message, or 30 seconds in the case of GPS L1/CA, to ensure
that the entire spoofed satellite ephemeris for each spoofed satellite could be
decoded. Fig.  \ref{fig:experimentalSetup} shows locations relevant to the
demonstration.  In the context of this paper, the physical location of the
transmitter (spoofer) is in Perth, Australia and the spoofed location sits atop
the former Aerospace Engineering building in Austin, Texas.  Note that this
spoofer could also be characterized as a meacon with a long delay from
reception to transmission. The goal is to geolocate the spoofer's position in
Perth. 

\subsection{Experimental Spoofer Geolocation with $\gamma$}

\begin{figure}[t]

    \begin{minipage}[b]{0.5\textwidth}
        \centering
        \includegraphics[width=\linewidth]{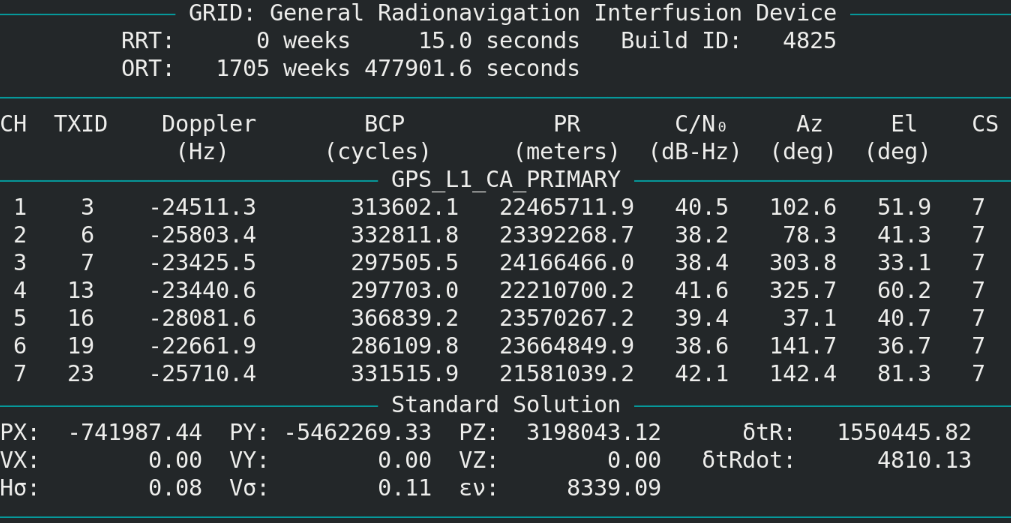}
    \end{minipage}
    
    \vspace{.75mm}
    
     \begin{minipage}[b]{0.5\textwidth}
        \centering
        \includegraphics[width=\linewidth]{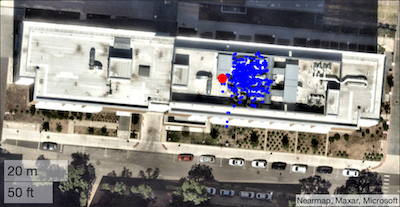}
    \end{minipage}
    \caption{Left: UT RNL's GRID receiver display when processing the
		spoofing signals.  Right: A scatter of GRID-derived position solutions. 
		The red dot is the spoofed position.  The 3D bias is 45.9 m, mostly concentrated in the vertical direction.  This error is attributed to the S-band carrier.}
	\label{fig:spoofedLocation}
\end{figure}

The transmitted spoofing signals captured in LEO were processed with the UT
RNL's GRID software-defined GNSS receiver \cite{clements2021bitpackingIonGnss,
nichols2022launch,pany2024historySdr}. Fig.  \ref{fig:spoofedLocation} shows
the PVT solution obtained by processing the pseudorange and Doppler
measurements of the spoofing signals.  The position solution is slightly biased
due to the code-carrier divergence caused by shifting the original L1-centered
signal to the S-band carrier \cite{clements2024DemoSpooferGeo}.  On GRID's
display, the 4,810 m/s clock drift (labeled $\delta \text{\tt tRdot}$) is
immediately noticeable.  Of course, no oscillator on a GNSS receiver would
experience a clock drift so extreme.

To coax GRID into properly processing the S-band spoofing signals, special
modifications to the receiver's configuration and PVT estimator had to be made.
Reconfiguring such parameters is trivial within GRID's software-defined
architecture.  The bandwidths of the receiver's delay lock loop (DLL) and PLL
were increased to maintain lock despite the code-carrier divergence introduced
by the S-band carrier.  The bandwidth of the DLL was set to 1.7 Hz and the
bandwidth of the PLL was set to 40 Hz, introducing more noise.   To minimize
spurious variations in $\hat{\gamma}(t)$, the receiver's dynamics model was set
to `static', consistent with an assumed static spoofed location.  The
receiver's innovations-based anomaly monitor was disabled to prevent rejection
of the PVT solution due to the unusually high estimated clock drift-rate.
Other considerations related to the S-band carrier are detailed by
\cite{clements2024DemoSpooferGeo}.

\begin{figure}[t]
	\centering
	\includegraphics[width=.98\linewidth]{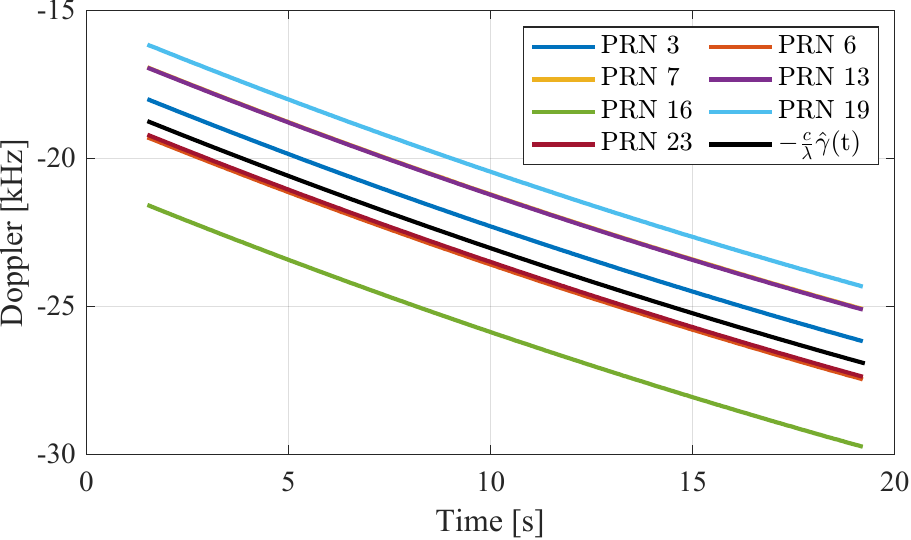}
	\caption{Measured Doppler time history of each received spoofing
		signal.  Also shown is the Doppler-equivalent time history $\hat{\gamma}(t)$ (black trace), which is used for geolocation.}
	\label{fig:Dopplers}
\end{figure}

A Doppler-equivalent time-history $\hat{\gamma}(t)$ over 17.75 seconds is shown
as the black trace in Fig.~\ref{fig:Dopplers} along with the raw measured
Doppler of each spoofing signal.  The GNSS receiver  allowed itself to be
spoofed and the true range-rate between the LEO-based receiver and the
terrestrial transmitter was lumped in the receiver's clock drift estimate as
explained in Section \ref{sec:conceptual}.  The measured Doppler time history
of each spoofing signal, as given in (\ref{eq:observedDoppler}), follows the
shape of $\hat{\gamma}(t)$ because the range-rate between the spoofer and
LEO-based receiver is dominant in all traces.  The deviation in the measured
Doppler time history of each spoofing signal from $\hat{\gamma}(t)$ is
$\tilde{f}_n(t)$, as presented earlier.

\begin{figure}[t]
	
	\begin{minipage}[b]{0.5\textwidth}
		\centering
		\includegraphics[width=.97\linewidth]{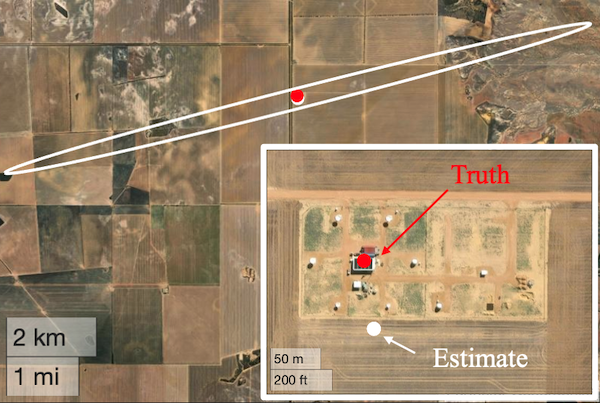}
		
		\vspace{6mm}
		
	\end{minipage}
	\begin{minipage}[b]{0.5\textwidth}
		\centering
		
		\includegraphics[width=.99\linewidth]{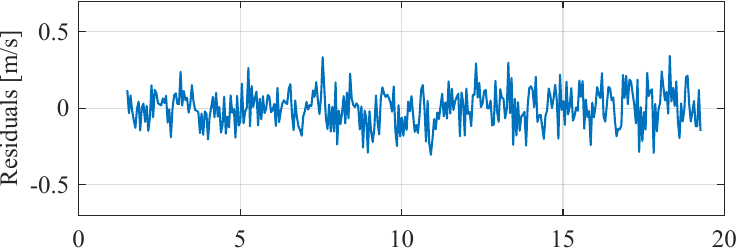}
		
		\vspace{2mm}
		
		\includegraphics[width=.99\linewidth]{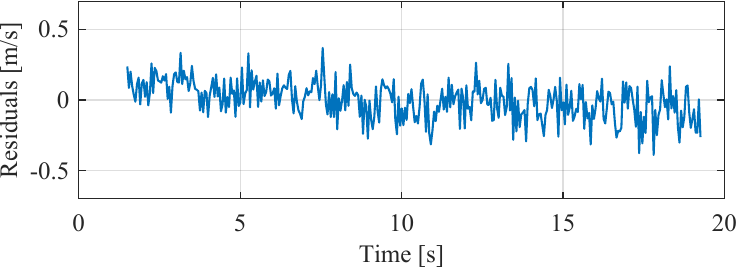}
		
	\end{minipage}
	
	\caption{Left: Final spoofer position estimate (white) based on $\hat{\gamma}(t)$.  Shown
		in red is the true spoofer location. The error of the final estimate is 68
		m. The true emitter is contained within the 95\% horizontal error ellipse, derived from (\ref{eq:Pxx}), which has a
		semi-major axis of 6.7 km. Right: Post-fit range-rate residuals of $\hat{\gamma}(t)$ time history with respect to the estimated spoofer position (top) and true spoofer position (bottom). The residuals with respect to the estimated position are unbiased and have a standard deviation of 0.12 m/s.}
	\label{fig:geofix}
\end{figure}

The time history of $\hat{\gamma}(t)$ was fed to the nonlinear least-squares
estimator described in Section \ref{sec:estimator}. The final position fix,
shown in Fig.~\ref{fig:geofix}, was within 68 meters of the true location.
Importantly, the true emitter position lay within the estimate's horizontal
95\% error ellipse.  For the measurement covariance matrix,  $\sigma_\text{a}$
was set to 0.15 m/s, and $\sigma_v$ was set to 0.0163 m/s, which is consistent
with the transmitter's TCXO.  The error ellipse's eccentricity is dictated by
the receiver-transmitter geometry.  Shown in Fig.  \ref{fig:geofix} are the
Doppler post-fit residuals, with respect to the estimated spoofer position and
the true spoofer position.  The residuals with respect to the estimated spoofer
position are zero-mean with a standard deviation of 0.12 m/s.  Such small and
unbiased residuals indicate that the estimator's model for $\hat{\gamma}(t)$ is
highly accurate. This experiment provides a validation of this paper's
geolocation technique.

\subsection{Experimental Spoofer Geolocation with GNSS Observables}

This paper's advocated technique requires a means of getting ephemerides and
clock models of the spoofed navigation satellites implied in the spoofing.  But
for cases in which the GNSS receiver onboard a LEO satellite cannot be
configured to produce a PVT solution from the spoofed signals, yet does produce
standard Doppler observables for each spoofed signal, traditional Doppler-based
geolocation as in \cite{murrian2021leo} can be applied to estimate the
spoofer's location.  Of course, as shown earlier, this will yield a biased
estimate of the spoofer's position because the time-varying frequency term
$\tilde{f}_n(t)$ is unmodeled.  However, if the spoofing signals induce a
static terrestrial location, the position bias due to the nonzero
$\tilde{f}_n(t)$ is small enough that the geolocation solution remains useful.

\begin{figure}[H]
	\centering
	\begin{minipage}[b]{0.49\textwidth}
		\centering
		\includegraphics[width=.9\linewidth]{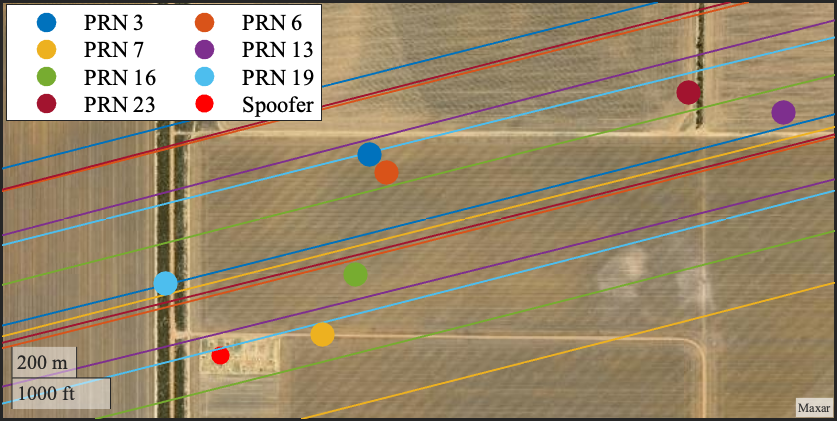}
		\begin{table}[H]
			\centering
			\scalebox{0.85}{
				\begin{tabular}{c | c c c c c c c} 
					\toprule
					PRN & 3 &  6 &  7 &  13 &  16 &  19 &  23 \\ \midrule
					Error [m] & 772 & 763 & 322 & 1,897 & 487 & 281 & 1,661 \\ 
					\bottomrule
				\end{tabular}
			}
		\end{table}
		\caption{Geolocation using the observed Doppler time history of each
			spoofed PRN.  Each individual spoofer position estimate is biased due to
			the unmodeled frequency component. }
		\label{fig:geofix_All}
	\end{minipage}
	\hspace{.01\textwidth}
	\begin{minipage}[b]{0.49\textwidth}
		\centering	
		\includegraphics[width=.99\linewidth]{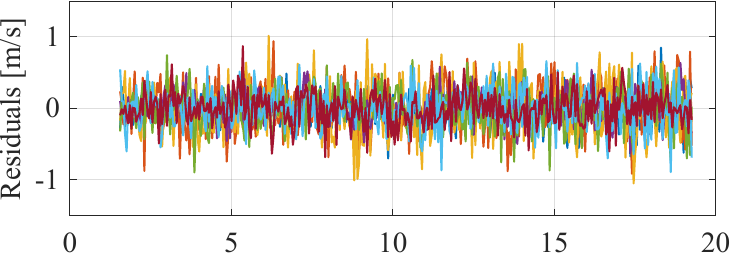}
		
		\vspace{3mm}
		
		\includegraphics[width=.99\linewidth]{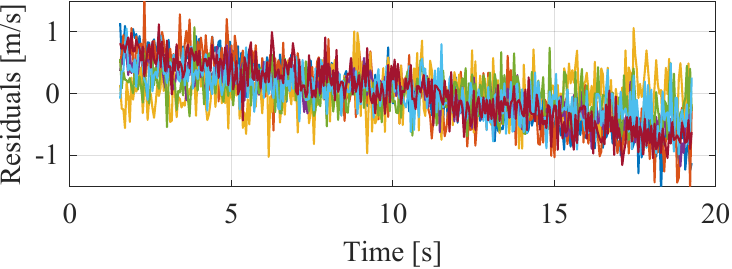}
		
		\vspace{5mm}
		
		\caption{Top: Range-rate residuals with respect to the estimated spoofer position.  Bottom: Range-rate residuals with respect to the true spoofer position.}
		\label{fig:geofix_All_residuals}
	\end{minipage}
\end{figure}

The position bias is relatively small because the Doppler time rate of change
between a stationary receiver on the surface of the Earth and a GNSS satellite
in medium Earth orbit is never more than 1 Hz/s, and typically much smaller.
Thus the range-rate between the LEO-based receiver and the physical spoofer is
the dominant term in $f_n(t)$.  Shown in Fig. \ref{fig:geofix_All} are the
biased position fixes and corresponding error ellipses when each $f_n(t)$ time
history is fed as measurements to the nonlinear least-squares estimator as
described in Section~\ref{sec:estimator}.  Only two of the seven 95\% error
ellipses contain the true spoofer position.  The spread of the spoofer position
estimates is relatively tight, with the maximum error being 1.9 km.  Depending
on the desired accuracy requirements, this level of accuracy may be sufficient.
Note that if the spoofer's induced trajectory were dynamic rather than static,
the spread of the geolocation estimates would be larger, as shown by
\cite{clements2022spoofergeo}.

Shown in Fig. \ref{fig:geofix_All_residuals} are the range-rate residuals with
respect to the estimated spoofer positions (top panel) and the true spoofer
position (bottom panel).  In the range-rate residuals with respect to the true
spoofer position, the time-varying frequency component is visible, especially
for PRNs 13 and 23, which also yield the final spoofer position estimates with
the largest amount of error.

\section{Spoofer Clock Instability Error Analysis} \label{sec:error_analysis}

This section analyzes how transmitter clock instability translates to
range-rate-based geolocation positioning error. It is important to characterize
such errors as they manifest in real-world applications.  In this section,
assume that $\delta \dot{t}_{\tilde{\text{r}}} = 0$ so that the effects of
actual---not induced---clock instability may be considered in isolation.  The
marginal contribution of transmitter clock instability to horizontal
positioning error scales directly with the transmitter oscillator quality,
specified by $h_{-2}$ in (\ref{eq:sig_v}).  This general result applies to any
clock quality and any capture geometry. 

As an example consider the capture scenario in
Section~\ref{sec:experimental_results} for a 20-second capture over Perth.
Table \ref{table:clock_quality} shows the contribution of transmitter clock
instability to the 95\% horizontal error ellipse semi-major and semi-minor axes
in the absence of all other error sources.  The orientation of the error
ellipse is determined by the capture geometry.  In general, the semi-major axis
lies in the cross-track direction of the satellite's motion, while the
semi-minor lies in the along-track direction.  Table \ref{table:clock_quality}
shows that single-satellite range-rate-based geolocation is sensitive to the
transmitter clock quality.  Thus, a spoofer could in theory use a low-quality
oscillator to degrade geolocation accuracy.  But its spoofing signals would
then more easily be detected by victim receivers, as will be discussed in the
next section, rendering it a less-effective spoofer.

The importance of correctly modeling $R$ is emphasized here using Monte Carlo
trials to compare two key metrics in geolocation: root mean square error (RMSE)
between the true and estimated spoofer position, and containment percentage.  

\begin{table}[H]
	\centering
	\begin{tabular}{lc  rr}
		\toprule
		Clock Quality & $h_{-2}$  & Semi-major [m] & Semi-minor [m]   \\ \midrule
		Low-quality TCXO & \num{3e-19} & 51,449 & 2,033 \\
		TCXO & \num{3e-21} & 5,145 & 203 \\
		Low-quality OCXO & \num{3e-23} & 514 & 20     \\
		OCXO & \num{3e-25} & 51 & 2 \\ \bottomrule
	\end{tabular}
	\caption{Theoretical marginal contribution of transmitter clock instability to
		the 95\% horizontal error ellipse for the capture scenario specified in
		Section~\ref{sec:experimental_results} in the absence of all other error sources. }
	\label{table:clock_quality}
\end{table}

\begin{figure}[t]

    \begin{minipage}[b]{0.49\textwidth}
        \centering
        \includegraphics[width=.99\linewidth]{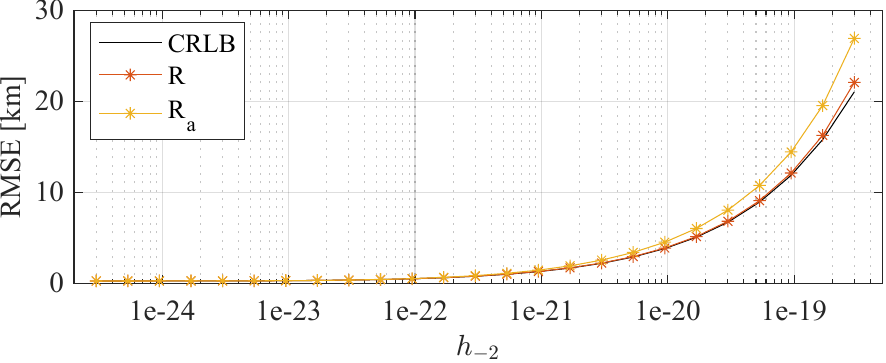}
    \end{minipage}
    \hspace{.02\textwidth}
    \begin{minipage}[b]{0.49\textwidth}
        \centering
        \includegraphics[width=.99\linewidth]{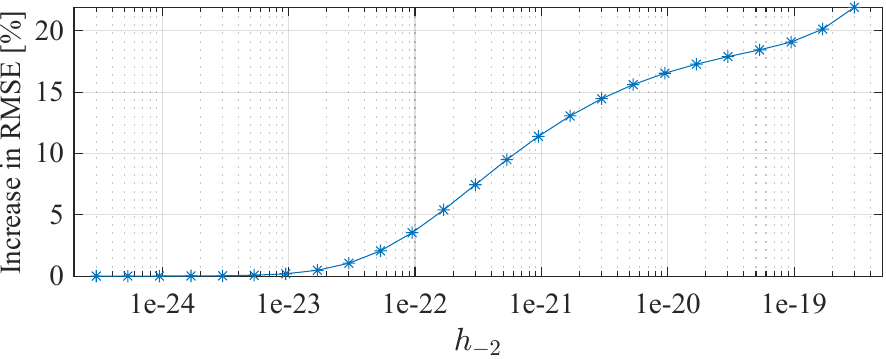}
    \end{minipage}
    
    \caption{Left: Monte Carlo sample RMSE as a function of $h_{-2}$ with the
          capture geometry specified in Section \ref{sec:experimental_results},
          for an estimator applying the correct ($R$) and incorrect
          ($R_\text{a}$) measurement covariance.  Right: Percentage increase in
          sample RMSE when $R_\text{a}$ is applied rather than $R$.}
	\label{fig:RMSE}
\end{figure}

For the RMSE comparison, the true range-rate time history for the 20 second
capture scenario specified in Section \ref{sec:experimental_results} was
computed.  For each Monte Carlo trial, both a realization of Gaussian random
walk consistent with a specified $h_{-2}$ and AWGN with $\sigma_a$ = 0.1~m/s
were added to the true range-rate.  The noisy range-rate measurements were
served to the nonlinear least-squares estimator with the correct measurement
covariance $R$ as specified in (\ref{eq:R}), and then with an incorrect
measurement covariance equal to $R_\text{a}$ (i.e., $R_\text{b}$ in
(\ref{eq:R}) was set to zero).  After the 10,000 Monte Carlo trials, the sample
RMSE was calculated for the sets of geolocation estimates corresponding to $R$
and $R_\text{a}$.  This was repeated with various $h_{-2}$ values
representative of a range of oscillators from low-quality TCXO to OCXO.  The
results are shown in Fig. \ref{fig:RMSE}.

One notes that the sample RMSE exhibited when using the correct measurement
covariance $R$ nearly achieves the CRLB. By contrast, erroneously modeling the
measurement noise as AWGN, as is the case when only $R_\text{a}$ is used,
ignores the time correlation introduced by the transmitter clock instability,
resulting in a greater-than 20\% increase in RMSE when the transmitter is
driven by a low-quality TCXO.  To be sure, the degradation in RMSE is only
noticeable for $h_{-2} > 3\times 10^{-23}$, corresponding to a low-quality OCXO
or worse. The increase in RMSE becomes more prominent when a low-quality
oscillator drives the transmitter because in this case the unmodeled Gaussian
random walk process is the dominant contributor to the measurement noise,
increasing the correlation between measurements.

Although taking $R_\text{a}$ alone as the measurement covariance is incorrect,
an \emph{unbiased} estimate is still achieved.  Nonetheless, the associated
estimated state error covariance becomes erroneously low.  Using the correct
measurement covariance produces an unbiased minimum-variance estimate with
properly sized state error covariance.

In addition to yielding a worse RMSE, using $R_\text{a}$ results in a
significantly worse containment percentage within the corresponding theoretical
95\% error ellipse.  Containment percentage is the percentage of trials in
which the true transmitter position lies within the theoretical 95\% error
ellipse centered at the estimated location.

A separate study of 10,000 Monte Carlo trials was conducted, again with the
capture geometry specified in Section \ref{sec:experimental_results}.  For each
trial, both a realization of AWGN with $\sigma_\text{a}$ = 0.1 m/s and a
Gaussian random walk consistent with a TCXO with $h_{-2} = 3\times 10^{-21}$
were added to the true range-rate.  When the correct measurement covariance $R$
was used, the corresponding theoretical 95\% error ellipse contained the
transmitter in 95.31\% of trials, as expected by a properly modeled estimator.
The area of this 95\% error ellipse was 3.47 km$^2$.

\begin{figure}[t]
	
	\begin{minipage}[b]{0.49\textwidth}
		\centering
		\includegraphics[width=.99\linewidth]{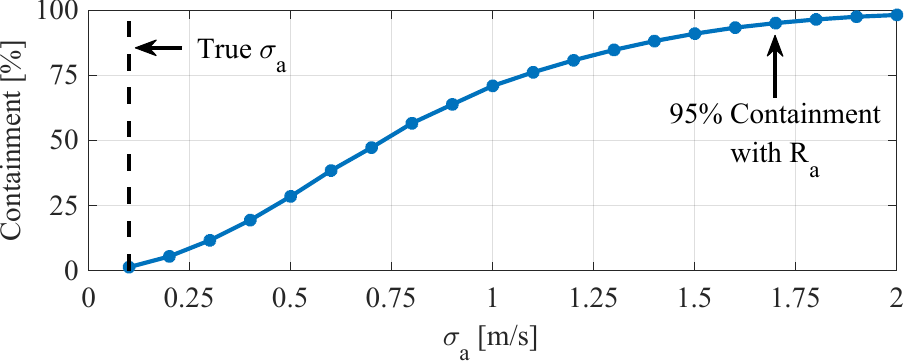}
	\end{minipage}
	\hspace{.02\textwidth}
	\begin{minipage}[b]{0.49\textwidth}
		\centering
		\includegraphics[width=.99\linewidth]{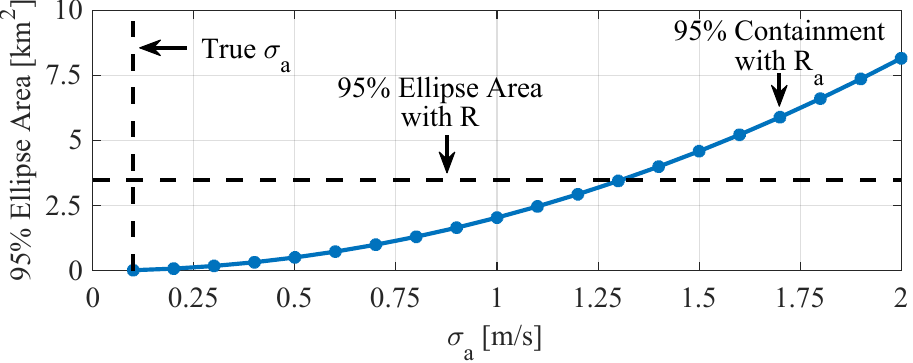}
	\end{minipage}
	\caption{Left: Monte Carlo containment percentage when the theoretical
		95\% error ellipse has been calculated with $R_\text{a}$ alone
		[setting $R_\text{b} = 0$ in (\ref{eq:R})], for various values of the
		underlying parameter $\sigma_\text{a}$.  Right: Area of the
		corresponding theoretical 95\% error ellipse as a function of
		$\sigma_\text{a}$. The horizontal line shows the area of the of the
		theoretical 95\% error ellipse with the correct full measurement
		covariance $R$. The vertical lines in both plots indicate the true
		value of $\sigma_\text{a}$ assumed in the Monte Carlo simulations.}
	\label{fig:Containment}
\end{figure}

By contrast, when $R_\text{b}$ was neglected and only $R_\text{a}$ was used,
there was significant degradation in the containment percentage.  For a case
with $R_\text{a}$ based on $\sigma_\text{a}$ = 0.1 m/s, the containment
percentage fell to 1.38\%.  Fig. \ref{fig:Containment} shows the containment
percentage for identical cases except with various different values of modeled
$\sigma_\text{a}$.  As one would expect, increasing the modeled
$\sigma_\text{a}$ improves containment percentage.  If $\sigma_\text{a}$ were
increased to 1.7 m/s, a 95\% containment percentage with $R_\text{a}$ is
achieved.  But this artificial inflation of $\sigma_\text{a}$ comes at the cost
of having a larger 95\% error ellipse.  Fig. \ref{fig:Containment} also shows
the area of the theoretical 95\% error ellipse for various values of
$\sigma_\text{a}$.  The area of the 95\% error ellipse for $\sigma_\text{a}$ =
1.7 m/s is 5.90 km$^2$, which is a 70\% increase in the 95\% error ellipse area
when compared using to the correct measurement covariance.  If
$\sigma_\text{a}$ were set to maintain the same 95\% error ellipse area as the
correct measurement covariance, a containment percentage of only 84.8\% is
achieved.

The spoofer's oscillator quality is typically unknown, which makes the
selection of the estimator's modeled $h_{-2}$ a design parameter.  The range of
plausible $h_{-2}$ is likely limited to TCXO quality or better, otherwise, the
transmitted spoofing signals are easily detectable by victims.  A multi-model
approach can be taken where the estimator of each model assumes a different
$h_{-2}$.  The convergence of the state estimate can be tested through a
goodness of fit test on the weighted sum of squared errors as presented by
\cite{blackman1999design}.

Properly modeling transmitter instability is thus essential in range-rate-based
geolocation so that the minimum-variance estimate is calculated and the
theoretical containment percentage is maintained. 

\section{Controlling Spoofing Detection While Degrading Geolocation Accuracy}

Researchers have developed formidable defenses against spoofing based on
receiver clock state monitoring \cite{jafarnia2013pvt, hwang2014receiver,
khalajmehrabadi2018real}.  A would-be spoofer has little flexibility to meddle
with the spoofed clock drift $\delta \dot{t}_{\tilde{\text{r}}}(t)$ if
intending to avoid detection by such defenses.  It follows that a stealthy
spoofer is scarsely able to purposefully degrade geolocation accuracy.

But consider a conspicuous spoofer---one willing to accept a potentially high
spoofing detection rate among affected receivers performing optimal time-based
spoofing detection.  In this case, the spoofer is allowed more flexibility to
manipulate $\delta \dot{t}_{\tilde{\text{r}}}(t)$ with the aim of either (1)
inflating victim receivers' timing error, or (2) confounding geolocation based
on this paper's technique.  This section derives and analyzes the attack
configuration that maximally increases geolocation error while maintaining a
specified detection rate among affected receivers implementing an optimal
receiver clock drift monitoring spoofing detection strategy.

\subsection{Optimal Spoofing Detection via Clock Drift Monitoring}

An optimal spoofing detection technique via receiver clock drift monitoring is
presented here.  Consider a time interval that spans $k~\in~\mathcal{K}~=~\{
1,2, ..., K \}$ uniformly sampled navigation epochs.  At the $k$th epoch, the
distribution of a GNSS receiver's measured clock drift $\delta
\dot{t}_{\text{u}}$ is modeled as
\begin{align}
  c\delta \dot{t}_{\text{u}} [k] \sim \mathcal{N}\left(c\delta \dot{t}_{\text{u}} [k-1],\, \sigma_{\text{u}}^{2} \right) \quad \text{where} \quad \sigma_{\text{u}}^{2}   = \sigma_{\text{m}}^{2}  + q 
\end{align}
is the steady-state measurement variance.  Here, $\sigma_{\text{m}}^{2}$ is the
component of the variance due to the measurement noise and clock dynamics
function, and $q$ is the process noise for $c\delta \dot{t}_{\text{u}}$, which
is related to the time between navigation epochs $\Delta t$ and the GNSS
receiver clock parameter $h_{-2}^\text{u}$ by \cite{brown2012introKf}
\begin{align}
  q & =2 \pi^2 h_{-2}^\text{u} \Delta t c^2
\end{align}
Let 
\begin{align}
  \eta_k = \frac{c\delta \dot{t}_{\text{u}} [k] - c\delta \dot{t}_{\text{u}} [k-1]}{\sigma_{\text{u}}} \sim \mathcal{N}\left(0, 1 \right)
  \label{eq:etai}
\end{align}
be the normalized increment in measured receiver clock drift at the $k$th
epoch.  Assume that increments are independent so that
$\E{\eta_k \eta_j}~=~\delta_{ij}~\text{for all}~ k,j~\in~\mathcal{K}$.

Optimal spoofing detection amounts to a hypothesis test that attempts to
distinguish the null hypothesis $H_0$ (receiver unaffected by spoofing) from
the alternative hypothesis $H_1$ (receiver captured by spoofing).  Note that
this section focuses solely on $\delta \dot{t}_{\tilde{\text{r}}}[k]$, the
spoofed clock drift increment, while assuming that the spoofer's transmitter
clock drift $\delta \dot{t}_\text{t}(t) = 0$, which is opposite the preceding
section's assumption.  Additionally, this analysis assumes a static GNSS
receiver performing detection so that the focus is on time-based spoofing
detection. Let
\begin{align}
  \mu_k = \frac{c\delta \dot{t}_{\tilde{\text{r}}}[k] - c\delta \dot{t}_{\tilde{\text{r}}}[k-1]}{\sigma_\text{u}}
  \label{eq:etai2}
\end{align}
be the normalized spoofed clock drift increment across one inter-epoch
interval, with initialization value $c\delta \dot{t}_{\tilde{\text{r}}}[0] = 0$
at $k = 0$, the moment when the spoofer captures the receiver.

Let $\theta_k$ represent the receiver's estimated clock drift increment at the
$k$th epoch under either hypothesis.  With the foregoing setup, this can be
modeled as
\begin{align}
	H_0&: \theta_k = \eta_k \; , \quad k \in \mathcal{K} 
	\\
	H_1&: \theta_k = \eta_k + \mu_k \;, \quad \mu_k \neq 0,  \quad k \in \mathcal{K} 
\end{align}

Under $H_1$, the value of $\mu_k$ is unknown to the receiver and belongs to the
set $\left( -\infty, 0\right) \cup \left( 0, \infty \right)$.  A uniformly most
powerful test does not exist for this hypothesis test because the critical
regions corresponding to $\mu_k < 0$ and $\mu_k > 0$ are different
\cite{vPoor1994dae}.  Instead, a locally most powerful (LMP) test is applied.
The LMP design problem is nearly the same as the Neyman-Pearson design problem,
so that the probability of detection is maximized while maintaining a fixed
probability of false alarm $P_\text{F}$.  For a single epoch, the detection
statistic $\Lambda^*(\theta_k)$ is 
\begin{align}
	\Lambda^*(\theta_k) = \theta_k^2 
\end{align}
and has the following distributions under $H_0$ and $H_1$
\begin{align}
  H_0&: \Lambda^*(\theta_k) \sim \chi^2_1 \\
  H_1&: \Lambda^*(\theta_k) \sim \chi^2_1\left(\lambda\right), \quad \lambda = \mu_k^2
\end{align}
where $\chi^2_n$ and $\chi^2_n(\lambda)$ denote, respectively, the chi-squared
and noncentral chi-squared distributions with $n$ degrees of freedom and
noncentrality parameter $\lambda$.

Consider detection based on data taken over a time interval that spans $K$
navigation epochs.  Let
$\vb{\theta}~=~\left[ \theta_1, \theta_2, ..., \theta_K \right]^\tr \in
\mathbb{R}^{K}$ and
$\vb{\mu}~=~\left[ \mu_1, \mu_2, ..., \mu_K \right]^\tr \in \mathbb{R}^{K}$.
The joint test statistic then becomes
\begin{align}
	\Lambda^*(\vb{\theta}) = \sum_{k\in\mathcal{K}}\theta_i^2 = \vb{\theta}^\tr\vb{\theta} 
\end{align}
with the following distributions under $H_0$ and $H_1$:
\begin{align}
	H_0&: \Lambda^*(\vb{\theta}) \sim \chi^2_K \\
	H_1&: \Lambda^*(\vb{\theta}) \sim \chi^2_K\left(\lambda\right), \quad \lambda = \sum_{k\in\mathcal{K}}\mu_i^2 = \vb{\mu}^\tr \vb{\mu}
\end{align}
An optimal-decision constant false alarm rate threshold $\nu^*$ for
$P_\text{F}$ can be calculated from
\begin{align}
	P_\text{F} &= P\left( \Lambda^*(\vb{\theta}) > \nu^* | H_0 \right) = 1 - F(\nu*; K)
\end{align}
where $F(\nu*; K)$ is the cumulative distribution function of $\chi^2_K$
evaluated at the detection threshold $\nu^*$.  The probability of detection is 
\begin{align}
  P_\text{D}(\vb{\mu}) &= P\left( \Lambda^*(\vb{\theta}) > \nu^* | H_1 \right) = 1 - F(\nu*; K, \lambda) \\
                       &= Q_{K/2}\left( \sqrt\lambda, \sqrt{\nu^*}\right)
                         \label{eq:PD}
\end{align}
where $F(\nu*; K, \lambda)$ is the cumulative distribution function of
$\chi^2_K\left(\lambda\right)$, and  $Q_m\left(\alpha, \beta\right)$ is the
Marcum Q function with $m~=~K/2$. The hypothesis test becomes 
\begin{align}
  \Lambda^*(\vb{\theta}) \mathop{\gtrless}_{H_0}^{H_1} \nu^*
  \label{eq:HT_vector}
\end{align}
The spoofer must optimize its attack configuration against this optimal
spoofing detection strategy.

\subsection{Expression for Geolocation Error}

One of the assumptions made when developing the estimator presented in Section
\ref{sec:estimator} was that $\delta \dot{t}_{\tilde{\text{r}}}$ is constant.
If instead $\delta \dot{t}_{\tilde{\text{r}}}(t)$ is time-varying, the
measurements $\hat{\gamma}[i]$ for all $i \in \mathcal{I}$ used for geolocation
become be perturbed, increasing geolocation error.  Let $\epsilon[i]$ represent
the unmodeled time-varying $c\delta \dot{t}_{\tilde{\text{r}}}[i]$ for all $i
\in \mathcal{I}$.  Then at the $i$th measurement epoch,
$c\hat{\gamma}[i]~=c\gamma[i]~+~\epsilon[i]$. Let the vector of measurement
perturbations over the capture interval be represented as $
\vb{\epsilon}~=~\left[\epsilon[1],~\epsilon[2],~...,~\epsilon[I]\right]^\tr \in
\mathbb{R}^I$, and let $\vb{\tilde{x}}~=~ [\tilde{e}, \tilde{n},
\tilde{b}]^\tr$ denote the geolocation estimation error in the east direction,
north direction, and frequency bias, where $\tilde{e}$ and $\tilde{n}$ are
defined in the East-North-Up (ENU) frame centered at the true spoofer position.
Let $\tilde{H} \in \mathbb{R}^{I\times 3} $ denote the measurement Jacobian
with respect to $\vb{\tilde{x}}$.  The error $\vb{\tilde{x}}$ can be calculated
as
\begin{align}
	\vb{\tilde{x}} = \left(\tilde{H}^\tr R^{-1}\tilde{H}\right)^{-1} \tilde{H}^\tr R^{-1}\vb{\epsilon} = B\vb{\epsilon}
\end{align}
The horizontal position error vector $\vb{e}_\text{h}$ is defined as 
\begin{align}
	\vb{e}_\text{h} =  \left[ \tilde{e}, \tilde{n} \right]^\tr
\end{align}
Let $\tilde{B}$ be the first two rows of $B$, and define the matrix
$A~\in~\mathbb{R}^{I \times I}$ as
\begin{align}
	A &= \tilde{B}^\tr \tilde{B}
\end{align}
The absolute horizontal positioning error $e_\text{h}$ due to the perturbation
$\vb{\epsilon}$ can then be computed as
\begin{align}
	e_\text{h} =  \sqrt{\vb{e}_\text{h}^\tr \vb{e}_\text{h}} = \sqrt{\vb{\epsilon}^\tr A \vb{\epsilon}}
\end{align}
Thus, the squared horizontal geolocation error $e_\text{h}^2$ is related to the
perturbation $\vb{\epsilon}$ by the quadratic form $\vb{\epsilon}^\tr A \vb{\epsilon}$.

The spoofer seeks the perturbation $\vb{\epsilon}$ that maximizes $e_\text{h}$
so that it can maximally degrade the accuracy of geolocation by a single sensor
platform performing range-rate-based geolocation via this paper's technique.
Suppose that $\vb{\epsilon}$ is subject to the constraint
$\|\vb{\epsilon}\| \leq \zeta$, which will be defined in the next section.  The
optimization problem then becomes
\begin{align}
	\vb{\epsilon}^* = \argmax_{\|\vb{\epsilon}\| \leq \zeta} \vb{\epsilon}^\tr A \vb{\epsilon} 
\end{align}
To solve this problem, $A$ is factorized as $A~=~Q D Q^\tr$, where $Q$ is
orthogonal and $D=\text{diag}(d_1, d_2, ..., d_I)$ is a diagonal matrix
composed of eigenvalues of $A$, which are all positive.  Assume that the
columns of $Q$ contain the unitary eigenvectors corresponding to eigenvalues
ordered such that $d_1~\geq~d_2~\geq~...~\geq~d_I$. Then
\begin{align}
	\vb{\epsilon}^\tr A \vb{\epsilon} = \vb{\epsilon}^\tr Q D Q^\tr \vb{\epsilon} = \vb{y}^\tr D \vb{y}
\end{align} 
where $\|\vb{\epsilon}\| = \|Q^\tr \vb{\epsilon}\| = \|\vb{y}\|$.  The value of
$\vb{y}$ that respects $\|\vb{y}\|~\leq~\zeta$ and maximizes $\vb{y}^\tr D \vb{y}$ 
is given by $\vb{y}^*~=~\left[ \zeta, 0, 0, ..., 0 \right]^\tr$.  Let
$\mathbf{v}^* \in \mathbb{R}^I$ denote the unitary eigenvector corresponding to
the largest eigenvalue of $A$.  The optimal $\vb{\epsilon}$ for this
optimization problem is then
\begin{align}
	\vb{\epsilon}^* = \pm \, Q \vb{y}^* = 	\pm \, \zeta \mathbf{v}^*
\end{align}

\subsection{Jointly Optimized Spoofer Clock Drift Selection}

Now that an optimal spoofing detector based on receiver clock drift has been
presented, and a perturbation $\vb{\epsilon}^*$ that maximizes horizontal
geolocation error subject to the constraint $\|\vb{\epsilon}^*\| < \zeta$ has
been defined, a spoofer can develop an attack configuration for $c\delta
\dot{t}_{\tilde{\text{r}}}(t)$ that maximizes $e_\text{h}$ while maintaining a
specified probability of detection.  It is assumed that the spoofer has perfect
knowledge of the LEO-based receiver's position and velocity, which is
representative of a worst-case scenario.

Let $c\vb{\delta \dot{t}}_{\tilde{\text{r}}} = c \left[ \delta
\dot{t}_{\tilde{\text{r}}}[1], \delta \dot{t}_{\tilde{\text{r}}}[2], ...,
\delta \dot{t}_{\tilde{\text{r}}}[I] \right]^\tr \in \mathbb{R}^I $ represent
the spoofer's discretized time-varying attack configuration for $\delta
\dot{t}_{\tilde{\text{r}}}(t)$.  Suppose the spoofer sets $c\vb{\delta
\dot{t}}_{\tilde{\text{r}}} = \vb{\epsilon}^*$.  Then the vector of spoofed
clock drift increments over $K = I-1$ navigation epochs is equivalent to
\begin{align}
	\vb{\mu} &= \frac{\zeta}{\sigma_\text{u}} C \mathbf{v}^* \in \mathbb{R}^K \quad \\\text{where} \quad \\	
    C &= 
	\begin{bmatrix}
		-1 & 1 & 0 & \dots & 0  \\
		0 & -1 & 1 & \ddots  & \vdots  \\
		\vdots & \ddots & \ddots & \ddots  & 0  \\
		0  & \dots & 0  & -1 & 1  
	\end{bmatrix} \in \mathbb{R}^{K \times I} \nonumber
\end{align}

The only task remaining for the spoofer is to determine the value of $\zeta$ so
that $\vb{\epsilon}^*$ can be scaled appropriately.  Suppose the spoofer is
willing to allow a detection probability $\bar{P}_\text{D}$ for the detection
test in (\ref{eq:HT_vector}).  Based on the parameters $\sigma_{\text{u}}$,
$I$, and $P_\text{F}$, the parameter $\zeta$ can be chosen to maintain an
expected probability of detection $\bar{P}_\text{D}$.  Given the functional
form of the probability of detection in (\ref{eq:PD}), $\zeta$ must satisfy the
equation
\begin{align}
	Q_{K/2}\left( \frac{\zeta}{\sigma_\text{u}} \| C\mathbf{v}^* \|, \sqrt{\nu^*}\right) = \bar{P}_\text{D}
\end{align}
Following this, the spoofed clock drift trajectory $c\vb{\delta
\dot{t}}_{\tilde{\text{r}}}^*$ that maximizes the geolocation error while
maintaining a specified probability of detection can be represented as
\begin{align}
  c\vb{\delta \dot{t}}_{\tilde{\text{r}}}^* = \pm \, \zeta \left(\mathbf{v}^* -
  \vb{1} \mathrm{v}^*[1] \right)
\end{align}
where $\vb{1}$ is the appropriately sized vector of all ones and
$\mathrm{v}^*[1]$ is the first element of $\mathbf{v}^*$.  Note that
subtracting $\vb{1} \mathrm{v}^*[1]$ ensures $c\vb{\delta
\dot{t}}_{\tilde{\text{r}}}[1] = 0$, consistent with initialization of the
spoofing attack. This subtraction does not change the optimization processes,
it only affects the estimated frequency bias $b_0$, which is merely a nuisance
parameter.

\begin{figure}[t]
	\centering
	
	
	\includegraphics[width=.99\linewidth]{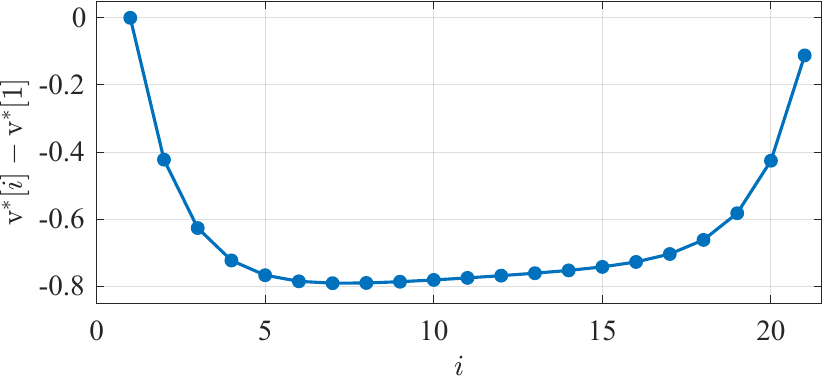}	
	
	\vspace{2.0mm}
	
	\caption{ The attack trajectory $\mathrm{v}[i]^* - \mathrm{v}^*[1]$ that
		maximizes $e_\text{h}$ for the LEO-based receiver geometry shown in
		Fig.~\ref{fig:experimentalSetup}.}
	\label{fig:worstCaseTrajectory}
	
\end{figure}

To illustrate the application of this analysis, consider the following example.
Suppose a spoofer wishes to choose $\vb{\delta \dot{t}}_{\tilde{\text{r}}}^*$
to maximally degrade geolocation by a LEO-based receiver capturing its signals
over 21 seconds with the geometry shown in Fig.~\ref{fig:experimentalSetup}.
Further suppose the LEO-based receiver computes measurements at 1 Hz, so that
$I$ = 21, and sets $R$ with $\sigma_\text{a} = 0.1$ m/s and $\sigma_v$
consistent with a TCXO.  The attack trajectory $\mathbf{v}^*~-~ \vb{1}
\mathrm{v}^*[1]$ that maximizes horizontal geolocation error is shown in
Fig.~\ref{fig:worstCaseTrajectory}.  It is interesting to note that the spoofer
allocates the greatest detection risk (largest increments) at the beginning and
end of the 21-second capture, while maintaining lower risk (smaller increments)
in the interim.

\begin{figure}[t]
	\centering
	\includegraphics[width=.99\linewidth]{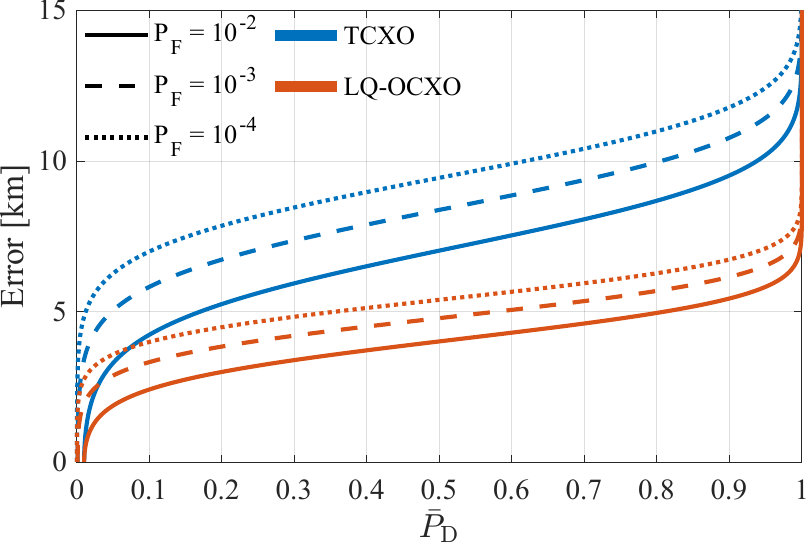}
	\caption{Worst-case geolocation error for a spoofer that optimizes
		$c\vb{\delta \dot{t}}_{\tilde{\text{r}}}$ for receivers performing 1-Hz
		spoofing detection tests with $\sigma_\text{m}$ = 0.05 m/s and for the
		LEO-based receiver geometry shown in Fig.~\ref{fig:experimentalSetup}.  The
		geolocation error is shown over a range of $\bar{P}_\text{D}$ for two
		representative victim receiver clock quality levels and three representative
		values of $P_\text{F}$. } 
	\label{fig:worstCaseError}
\end{figure}

Now assume that spoofing-affected receivers are performing navigation solutions
once per second with $\sigma_\text{m}$~=~0.05 m/s.  Shown in
Fig.~\ref{fig:worstCaseError} is the maximum horizontal geolocation error given
a triad of $\bar{P}_\text{D}$, $P_\text{F}$, and affected receiver clock
quality.  For example, if the spoofer accepts a detection rate of
$\bar{P}_\text{D}$~=~0.5 by receivers equipped with a TCXO having their
spoofing detector set with $P_\text{F}~=~10^{-3}$, the maximum $e_\text{h}$ due
to $c\vb{\delta \dot{t}}_{\tilde{\text{r}}}^*$ is 8.4 km.

To give the reader an idea of how capture geometry affects the maximum
horizontal geolocation error, consider the same scenario, but with a 21-second
detection-and-geolocation segment beginning 30 seconds earlier.  This capture
geometry is more favorable for geolocation.  It results in a maximum horizontal
geolocation error of 2.2 km.  On the other hand, consider a 21-second segment
beginning 30 seconds after the original.  This capture geometry is worse for
geolocation.  It results in a maximum horizontal geolocation error of 23.5 km.
It is important to note that this worst-case error is not a limitation of this
paper's technique, but a limit of single-satellite range-rate-based geolocation
of GNSS spoofers in general. And it should be remembered that the foregoing
analysis is for a worst-case situation in which the spoofer knows the LEO-based
receiver's position and velocity time history.

\section{Conclusion}

This paper presented a single-satellite, single-pass technique for locating
GNSS spoofers from LEO.  The technique was validated in a controlled experiment
in partnership with Spire Global in which a LEO-based receiver captured GNSS
spoofing signals transmitted from a ground station.  An analytic expression for
how actual transmitter clock instability degrades the geolocation solution was
derived.  Finally, geolocation positioning error as a function of worst-case
spoofed clock behavior subject to a constraint on probability of detection was
investigated.

\section*{Acknowledgments} 
This work was supported by the U.S. Department of Transportation under Grant
69A3552348327 for the CARMEN+ University Transportation Center, and by
affiliates of the 6G@UT center within the Wireless Networking and
Communications Group at The University of Texas at Austin.

\bibliographystyle{ieeetr} 
\bibliography{sample_ieee.bib}
\end{document}